\documentstyle[12pt]{article}
\global\arraycolsep=2pt  
\begin{document}
\begin{titlepage}
 \vspace{0.3cm}
\begin{center}
\Large\bf  
Lessons from $QCD_2 (N\rightarrow\infty)$:\\
  Vacuum structure, Asymptotic Series, Instantons
       and all that...    \end{center}

\vspace {0.3cm}

 \begin{center}    {\bf Ariel R. Zhitnitsky}\footnote{
  e-mail address:arz@physics.ubc.ca }
 \end{center}
\begin{center}
{\it Physics Department, University of British Columbia,
6224 Agricultural  Road, Vancouver, BC V6T 1Z1, Canada}
\end{center}
\begin{center}
\vspace{0.3cm}

\end{center}
\begin{abstract}
 
 We discuss two dimensional $QCD (N_c\rightarrow\infty)$ 
with fermions in the fundamental
  as well as   adjoint representation. 
We find  factorial growth $\sim (g^2N_c\pi)^{2k}\frac{(2k)!(-1)^{k-1}}{(2
\pi)^{2k}}$ in the coefficients of the
large order perturbative expansion. 
 We argue that this behavior is related to   
classical solutions
of the theory, instantons, thus it has nonperturbative origin.
Phenomenologically such a growth is related to highly 
 excited states in the spectrum. 
We also analyze the heavy-light quark system $Q\bar{q}$
within operator product expansion 
(which it turns out to be an asymptotic series).
Some vacuum condensates $\langle\bar{q}(x_{\mu}D_{\mu})^{2n}q\rangle 
 \sim (x^2)^n\cdot n!$ which are responsible for this 
factorial growth are also discussed.

We formulate some general puzzles which are not specific
for 2D physics, but are inevitable features  of any asymptotic expansion.
We resolve these apparent puzzles within $QCD_2$ and we speculate that
analogous puzzles might occur in real 4-dimensional QCD as well.

\end{abstract}
\end{titlepage} 
\vskip 0.3cm
\noindent

\setcounter{page}{1}

\section{ Introduction}

\indent\indent 
The problem of large -order behavior of the perturbative theory
attracted a renewed attention recently.
 One of the motivating factors is a common wisdom that the corresponding
asymptotic behavior is related somehow to a very deep physics. 
This is the area where perturbative and nonperturbative
physics strongly interfere. An understanding of this interplay
may shed the light on the nature of the nonperturbative vacuum structure
in general and the origin of vacuum condensates in particular.

With these general remarks in mind we would like to analyze
these problems   
in solvable two-dimensional $QCD (N\rightarrow\infty)$,
\cite{Hooft}-\cite{Kent}. We would like  to test   assumptions, hypothesis
and interpretations,   made in 4-dimensional field theory,
within toy two-dimensional $QCD_2 (N\rightarrow\infty)$,
where we expect confinement and many other properties
inherent to real QCD.
 Additionally, we extend the analysis to 
QCD with adjoint matter \cite{Klebanov,Kutasov}. As is known, 
in this theory, the pair creation is not suppressed even in the
large $N_c$ limit, and thus, this model can mimic an exponentially
 growing density of states with large mass $\rho(m)\sim\exp m$.
In this case no exact solution is  available, but we argue
that   general methods such as dispersion relations, 
duality and  unitarity  can provide all information we need
about spectrum for the calculation of large order behavior.
  
Why are we so conscious about the large order behavior?
We see at least a  few theoretical and phenomenological reasons for that.
Let us start from the pure theoretical reasons.
One may think
that the  crucial question in this case is whether the
perturbative  series is Borel
summable or not.
 
Contrary to the common belief, we do not think that
the issue of Borel summability (or its loss )  is the fundamental
  point. In particular, let us mention
an example 
of the principal chiral field theory at large $N$ \cite{Fateev}.
In this case, the explicit solution  
as well as  the coefficients of the perturbative
expansion can be calculated. These coefficients grow factorially with
the order and the series is non-Borel summable, but
 nevertheless, 
the physical observables are perfectly exist for any finite
coupling constant. The exact result can be recovered by
special prescription which uses a non-trivial procedure of
 analytical continuation (which might be a good example
for other asymptotically free theories).

The second important theoretical issue can be formulated as follows:
Because of dimensionality of the coupling constant in $QCD_2$
the perturbative expansion $\sim \sum (g^2)^nc_n$ and 
the operator product expansion (OPE)  for 
a correlation function $\sim\sum(\frac{g^2}{Q^2})^nc_n$ are 
{\it one and the same} expansion. From this simple 
observation we   learn that the OPE is an asymptotic series.
Thus, many interesting questions 
arise: \\

a)What kind of vacuum condensates are responsible for such a 
behavior?
 
b)Do we extract the actual condensates from the OPE 
or only effective ones?

c)
What kind of vacuum  configurations are
responsible for such $n!$ growth?

d)Do these configurations saturate the vacuum condensates?
 
 What is more important, sometimes
these questions (and many others) can be answered. We expect,
  will argue,
that the analogous phenomena might occur in real
four dimensional QCD, thus these questions have not 
 only pure academic interest.

Phenomenologically, there are issues which are even more interesting 
and give much more freedom for speculation.
First of all, let us recall that the reason for   interest in 
the large
order behavior  is related to the   factorial $n!\simeq n^n$
growth of the perturbative coefficients. This growth can be considered
 via the dispersion relation 
and is commonly interpreted as a
reflection of the divergence of the multiparticle 
cross section with large number of particles 
  and energy $\simeq n$ (see discussions in \cite{Zakharov}).  
The naive interpretation would be the violation of unitarity 
\footnote{In the physical theory, the unitarity is
 preserved, of course. The physical question is:
 what can stop this growth?}.
We will show however  that in $QCD_2$  while we have a factorial growth
of coefficients,  this growth has nothing to do with
multiparticle production since at large $N$ the pair creation is suppressed
by factor $1/N$. Rather, this growth  is related to the highly excited two-particle
meson states. Another phenomenological issue looks mysterious:
vacuum condensates extracted from the
spectrum might be quite different from
the actual magnitude of condensates.

\section{ t'Hooft model}

\indent\indent Let us start from the  analysis of   
two dimensional QCD with  fermions in the  fundamental representation 
 -the 't Hooft model. It is completely solvable in the 
limit where the 
number of colors $N\rightarrow \infty$ \cite{Hooft}. 
The Bethe-Salpeter equation for mesonic bound states was solved in 
\cite{Hooft}
yielding a spectrum whose states lie asymptotically on a single 
"Regge trajectory".
We want to point out that many general questions in this
 model can be answered
without solving an equation, but using such powerful methods as
dispersion relations, duality and unitarity.
In particular, in the weak coupling regime,
\begin{eqnarray}
\label{1}
 g^2N\sim const. ~~~N\rightarrow\infty,~~~m_q
\gg g\sim\frac{1}{\sqrt{N}} 
\end{eqnarray}
the chiral $\langle\bar{\psi}\psi\rangle $ and gluon condensates 
$\langle G_{\mu\nu}^2\rangle $ can be calculated {\it exactly}, see below.
Additionally, few low-energy theorems can be tested and obtained  
result
imply that there are no other states in addition to those
found by 't Hooft. 
In other words,   the dispersion 
and duality relations
  would indicate missing states. 

 Here, the entire spectrum is discrete 
and is classified by the integer  $n$.  
 The model we shall consider consists of quark in fundamental
representation interacting via an $SU(N)$ color gauge group. We follow
the notation of ref.\cite{Callan} and present the 't Hooft equation
\cite{Hooft} in the following form
\begin{eqnarray}
\label{2}
m_n^2\phi_n(x) =\frac{m_q^2}{x(1-x)}\phi_n(x) 
-m_0^2P\int dy\frac{\phi_n(y)}{(x-y)^2},
\end{eqnarray}
where symbol $P$ notes as the principal value of the integral, and 
$0<x<1$ is the  fraction of the total  momentum
of the bound state  carried  by quark $q$ 
with mass $m_q$.
The quantity 
$m_0^2\equiv \frac{g^2N}{\pi}$ is the basic mass scale in the theory
and the index  $n$ classifies the ordering  number
 of the bound states $|n,p\rangle$
with total momentum $p_{\mu}$. 

The same wave function 
can be expressed in terms of the following matrix element 
\cite{Brower1}:
\begin{eqnarray}
\label{3}
\phi_n(x)=\sqrt{\frac{N}{\pi}}\int d y_+e^{-iy_+(1-2x)p_-}
\langle 0|\bar{q}(-y)q(y)|n,p\rangle|_{y_-=0}.
\end{eqnarray}
Let us note that the matrix element on the right   
is written in the light cone gauge $A_-=0$; to restore
the  manifest  gauge invariance one can insert
the standard exponential factor $e^{ig\int A_-dy_+}$
into the  formula (\ref{3}).

Let us review some important properties of  equation (\ref{2}).
The entire spectrum is discrete 
and classified by the integer number $n$. The wave functions $\phi_n(x)$
are orthogonal, complete and obey the following boundary conditions
\begin{eqnarray}
\label{4}
\phi_n(x)\rightarrow [x(1-x)]^{\beta},~~~x\rightarrow 0,
~~x\rightarrow 1,~~~
\pi\beta\cot(\pi\beta)=1-\frac{m_q^2}{m_0^2}.
\end{eqnarray}
For large   $n$ the spectrum is linear
\begin{eqnarray}
\label{5}
m_n^2\simeq \pi^2m_0^2 n,~~\phi_n(x)\simeq\sqrt{2}\sin(\pi n x)
\end{eqnarray} 
and does not depend on mass of the quark.
More importantly, in the chiral limit ($m_q\rightarrow 0$)
the lowest level (we call it $\pi$ meson) tends to zero
($m_{\pi}^2\sim m_q$) and one could expect a nonzero
magnitude for the chiral  condensate. 
Thus, we have  come    to the very important connection
between  spectrum 
and vacuum structure.  

As the vacuum  of the model is a very important issue
for the   following analysis, we would like to recall some
results   with an explanation of the 
 general methods which
have been used to derive them.

We define the chiral condensate in the current algebra terms  
as follows:
\begin{eqnarray}
\label{6}
0= \lim_{p_{\mu}\rightarrow 0} i\int d^2x e^{ipx}\partial_{\mu}
\langle 0|T\{\bar{q}\gamma_{\mu}\gamma_{5}q(x),~
\bar{q}\gamma_{5}q(0)\}|0\rangle =   \\   \nonumber
 2i \langle 0|\bar{q}q|0\rangle + 2m_q \cdot
 \langle 0|T\{\bar{q}i\gamma_{5}q(x),~
\bar{q}i\gamma_{5}q(0)\}|0\rangle .
\end{eqnarray}
As we have already mentioned, the only states  of
 't Hooft's solution are the    quark-antiquark bound  states,
thus they must saturate the dispersion relation. Upon inserting
this complete set of mesons to the (\ref{6}) one   obtains:
\begin{eqnarray}
\label{7}
 \langle 0|\bar{q}q|0\rangle =-m_q\sum_{n}\frac{N}{\pi}
\frac{f_n^2\pi^2m_0^2}{m_n^2},
\end{eqnarray}
where $f_n$ is defined in terms of the following matrix elements
\begin{eqnarray} 
\label{8}
 \langle 0|\bar{q}i\gamma_{5}q  |n\rangle
=\sqrt{\frac{N}{\pi}}\pi m_0f_n 
 ,~~f_n\pi m_0=\frac{m_q}{2}\cdot\int_0^1\frac{\phi_n(x)}{x(1-x)}dx
\end{eqnarray}
In the chiral limit the only   state which  
 can contribute
to the formula (\ref{7}) is the $\pi$ meson.
 Its matrix element can be calculated exactly
and we end up with the following expression for the chiral condensate
in the $m_q\rightarrow 0$ limit \cite{ARZ}:
\begin{eqnarray} 
\label{9}
 \langle 0|\bar{q}q|0\rangle =-N\frac{m_0}{\sqrt{12}},~~m_0^2=\frac{g^2N}{\pi},
~~f_{n=0}=\frac{1}{\sqrt{3}},~~m_{\pi}^2=m_q\cdot\frac{2m_0}{\sqrt{3}}.
\end{eqnarray}
The result is 
confirmed by numerical \cite{Ming1},\cite{Ming2} and 
independent analytical
calculations\cite{Lenz}. Moreover, the method has been  generalized for 
the nonzero quark mass and the corresponding explicit formula
for the chiral  condensate $\langle \bar{q}q\rangle$
with arbitrary $m_q$ has been obtained
\cite{Mattis}.

As was expected, we find that $ \langle 0|\bar{q}q|0\rangle \sim N$.
Besides that, as we have already noticed in \cite{ARZ}, if we put
$m_q=0$ from the very beginning, then $ \langle 0|\bar{q}q|0\rangle =0$. 
This  corresponds to the different regime when $m_q\ll g\sim1/\sqrt{N}$,
when nonplanar diagrams come into the game as  we will discuss  later.
The last remark is the observation that the entire nonzero answer
for the condensate comes from the infrared region of the 
integration in eq.(\ref{8}): $x\sim 0, x\sim1$ which corresponds to the
situation when one of the quarks carries all the momentum and the 
second one is at rest.

The sum (\ref{7}) can be calculated exactly
for arbitrary $m_q$ \cite{Mattis}. The crucial point is that
for arbitrary $m_q$
the nonzero contribution comes from the highly excited states ($
n\gg 1$) only.
The properties of these states are well-known:
\begin{eqnarray}
\label{10}
f_n^2\rightarrow 1, ~~~~~~m_n^2\rightarrow\pi^2m_0^2\cdot n,
~~~~n\gg 1,
\end{eqnarray}
and thus the sum (\ref{7}) can be explicitly evaluated with the result
\cite{Mattis}:
\begin{eqnarray}
\label{11}
 \langle 0|\bar{q}q|0\rangle=\frac{m_qN}{2\pi}
\{\log(\pi\alpha)-1-\gamma_E+(1-\frac{1}{\alpha})[I(\alpha)-
\alpha I(\alpha)-\log 4]\},
\end{eqnarray}
where $\alpha=\frac{m_0^2}{m_q^2},~~\gamma_E=0.5772..$
is Euler's constant and
$$ I(\alpha)=\int_0^{\infty}\frac{dy}{y^2}\frac{1-\frac{y}{\sinh y \cosh y}}
{[\alpha(y \coth y -1) +1]}.$$
This result is exact for large $N$  and arbitrary quark mass
within the 't Hooft regime, i.e.
$m_q\gg g\sim1/\sqrt{N}$ (\ref{1}). In the 
limit $\alpha\rightarrow\infty$,  it reduces to the eq.(\ref{9})as
 it should.

The last condition ($m_q\gg g $ ) which has to be
 satisfied for the 't Hooft solution to be valid,
requires some additional explanation.
Roughly speaking, nonplanar diagrams may contain
a factor $\sim m_q^{-1}$ which at $m_q=0$ blows up and the theory 
changes
completely. The concept of the proof that there exists a factor
 $\sim m_q^{-1}$ in nonplanar diagrams is the following.

Let us consider the correlation function for $p\rightarrow 0$
\begin{eqnarray}
\label{12}
  i\int d^2x e^{ipx} 
 \langle 0|T\{\bar{q} q(x),~
\bar{q} q(0)\}|0\rangle =P(p^2 )
\end{eqnarray}
 The 't Hooft solution suggests that only 
planar graphs   are taken into account and, 
consequently, the spectral
density contains only   contributions from one meson
states for which     $P_{planar}\sim N$.
At the same time in the chiral limit, we can   calculate
the  two-pion contribution  exactly!
This contribution    
 is {\bf not accounted for }
in deriving (\ref{2}). Of course, the two-pion
contribution is suppressed by a factor $1/N$. However,
it contains a term $\sim\frac{m_0^2}{m_{\pi}^2}$
which tends to infinity for $m_q\rightarrow 0$.
The presence of the factor $\sim m_q^{-1}$
in nonplanar diagrams leads to the aforementioned constraint 
on $m_q$.

Now let us explicitly demonstrate the existence
of the term $\sim m_q^{-1}$ for the two-pion contribution.
In order to do so, we write down a
 dispersion relation for $P$:
\begin{eqnarray}
\label{13}
P(0)=\frac{1}{\pi}\int_{4m_{\pi}^2}^{\infty}\frac{ds}{s}
 ImP(s) ,
\end{eqnarray}
where $ImP(s)$ is the physical spectral density.
The $\pi\pi$ contribution is fixed uniquely by (\ref{9})
 because of the special
role of pions \cite{ARZ}:
\begin{eqnarray}
\label{14}
\langle \pi\pi|\bar{q}q|0\rangle|_{p\rightarrow0}=\frac{m_0\pi}{\sqrt{3}},~~
\frac{1}{\pi}
ImP^{\pi\pi}(s)=\frac{m_0^2\pi^2}{6}\frac{1}{\sqrt{s(s-4m_{\pi}^2)}},
\\     \nonumber
P^{\pi\pi}(0)=\frac{m_0^2\pi^2}{6}\int_{4m_{\pi}^2}^{\infty}
\frac{ds}{s\sqrt{s(s-4m_{\pi}^2)}}=
\frac{m_0^2\pi^2}{12 m_{\pi}^2}\sim\frac{1}{m_q}.
\end{eqnarray}
It is clear that the only cause for 
a singular $\sim 1/m_q$ behavior is the 
finiteness of the pion matrix elements at zero momentum.
At the same time this contribution does not contain
the  large factor $N$ which accompanies a one meson 
contribution to the same
correlator. 
To suppress these nonplanar  diagrams we  
require $N\gg\frac{m_0^2}{m_{\pi}^2}$.
Thus, we   expect that some kind of phase transition may occur
in the region $m_q\sim g$, which would couse a complete
restructuring of the theory.

The last subject we would like to discuss in this section
is the strict Coleman theorem  \cite{Coleman}
which states that a continuous symmetry cannot be broken
spontaneously in a two dimensional theory.
As we discussed earlier \cite{ARZ} we expect that as in the 
$SU(N\rightarrow\infty)$ Thirring model 
(where the chiral symmetry
is ``almost" spontaneously broken \cite{Witten}),
the Berezinski-Kosterlitz-Thouless
(BKT )  effect \cite{BKT} operates in regime (\ref{1}).
This fact also confirms the 't Hooft spectrum:
states with opposite $P$ parity are not degenerate
in mass and there is an ``almost" Goldstone boson
with $m_{\pi}^2\sim m_q+1/N$.

To be more specific, one can show \cite{ARZ} that in 
$QCD_2 (N\rightarrow\infty)$ 
the behavior of the proper
two-point correlation function 
is as follows:
\begin{eqnarray}
\label{15}
    \langle 0|T\{\bar{q}_{L} q_{R}(x),~
\bar{q}_{R} q_{L}(0)\}|0\rangle \sim x^{-\frac{1}{N}}.
\end{eqnarray}
Such a behavior together with cluster  
property as $x\rightarrow\infty$ implies 
the existence of the condensate at $N=\infty$
in a full agreement with our previous discussion.
At the same time, for any finite but large  $N$, the correlator
falls off very slowly demonstrating the BKT-behavior
with no signs of contradiction to the Coleman theorem.

Having these general remarks on $QCD_2(N)$
in mind, we turn into our main subject.
   \section {Large order behavior in $QCD_2(N=\infty)$.}
\subsection{'t Hooft model.}

Let us   consider the
asymptotic limit $ Q^2=-q^2\rightarrow\infty $
of the two-point correlation function \cite{Callan}, \cite{ARZ}:
\begin{equation}
\label{1a}
i\int dx e^{iqx} \langle 0|T \{ \bar{q}i \gamma_5q(x),
\bar{q}i \gamma_5q(0) \} |0 \rangle = P(Q^2).
\end{equation}
It is clear, that the large $Q^2$ behavior of $P(Q^2)$ is
 governed by
the free, massless theory, where
\begin{equation}
\label{2a}
P(Q^2\rightarrow\infty)=-\frac{N_{c}}{2\pi}\ln Q^2.
\end{equation}  
At the same time the dispersion relations state that
\begin{equation}
\label{3a}
P(Q^2)=\frac{N_{c}m_0^2\pi^2}{\pi}\sum_{n=0,2,4,...}\frac{f_n^2}{Q^2+m_n^2}
\end{equation}  
and the  sum is over    states with   even   $n$  because
 we are considering the pseudoscalar currents. 
 Here residues  $f_n$ are defined as follows
\begin{equation}
\label{4a}
 \langle 0|\bar{q}i \gamma_5q |n \rangle=\sqrt{\frac{N_{c}m_0^2\pi^2}{\pi}}f_n  
,~~~n=0,2,4,.. . 
\end{equation}  

Bearing in mind that for large  
$n,~~f_n^2\rightarrow 1$
and $m_n^2\rightarrow m_0^2\pi^2n$, we recover the
asymptotic result (\ref{2a}). We can reverse arguments by saying that  
in order
to reproduce $\ln Q^2$ dependence in 
the dispersion relation (\ref{2a}), the residues $f_n^2$  
must
approach the constant $\sim (m_{n+1}^2-m_n^2)$ for large $n$.

Now consider a  $Q^{-2k}$ expansion 
for the correlator (\ref{3a})
in order to find the coefficients $c_k$ of this series at large $k$
\begin{equation}
\label{c_k}
P(Q^2)\sim \sum_k c_{2k}g^{2k}\sim \sum_kc_{2k} (\frac{g^2N_c}{Q^2})^{k} .
\end{equation} 
As we mentioned earlier, in two dimensions the perturbative expansion
$c_k(g^2N)^k$ and the $1/(Q^2)^k$ -expansion  coincide.

Now, if we knew $f_n$ and $m_n$ for arbitrary $n$ we could
 calculate the sum (\ref{3a}) precisely, and thus, we would find
the coefficients $c_k$ from (\ref{c_k}). Unfortunately,
we do not know them.
 However,
the key observation is as follows: in spite of the fact 
that we do not know an analytical expression
for $f_n$ and $m_n$ for arbitrary $n$ we still  can calculate 
the leading behavior of  $c_k$.
  The reason for that is related to the fact that 
the only asymptotics of residues
$f_n=1,~n\rightarrow\infty$ and masses $m_n^2=
m_0^2\pi^2n,~n\rightarrow\infty$ are essential;
The corrections to $f_n=1+0(1/n),~ m_n^2=m_0^2\pi^2n+0(1),~
~n\rightarrow\infty$ might change the preasymptotic behavior of     
$c_k,~k\rightarrow\infty$, but can not change
the factorial behavior $(k)!$, found
below.
 Using the asymptotic expressions 
for $f_n$ and $m_n$ we find that $P(Q^2)$ is expressed
in terms of transcendental function 
$\Psi(z)=\frac{\Gamma^{\prime}(z)}{\Gamma(z)}$ where
$z=\frac{ Q^2}{m_0^2\pi^2}$. However we can trust
only in the leading terms of  the corresponding formula:
\begin{eqnarray} 
\label{5a}
P(Q^2)-P(0)=\frac{N_c}{2\pi}
\sum(\frac{1}{n+z}-\frac{1}{n})=-\frac{N_c}{2\pi}
\left[\ln z + \gamma_E+\frac{1}{2z} -
\sum_{k=1}^{\infty}
 \frac{B_{2k}}{2k}\cdot\frac{1}{z^{2k}}\right],~ \nonumber
\end{eqnarray}
\begin{eqnarray}
z= \frac{Q^2}{m_0^2\pi^2}~~~
B_{2k}=\frac{(2k)!2(-1)^{k-1}}{(2\pi)^{2k}}
\end{eqnarray}
where $B_{2k}$ is asymptotic expression for the Bernoulli numbers.

A few comments are in order.
First of all, we have explicitly demonstrated that the 
coefficients $c_{2k}$ in the operator
expansion $\sum c_{2k}/Q^{2k}$ are
 factorial divergent in high orders, $c_{2k}\sim(2k)!$, so the
expansion is asymptotic in full agreement with the general
arguments of ref. \cite{Shifman}.

Additionally we  note that the only 
even powers $g^2/Q^2$ are essential in the expansion ( generally speaking,
 arbitrary powers of $g^2/Q^2$
could contribute). 
Nonleading terms in $f_n^2$ and $m_n^2$ might contribute to the odd
powers $g^2/Q^2$. 

From the physical point of view this factorial 
behavior is related to highly excited states with
excitation number $2n_0\sim Q^2/m_0^2$, and not with
multiple production as one might naively   expect. Indeed
let us consider $Q^2/m_0^2\sim z\leq 2n_0$
in the expansion $ \sum\frac{1}{2n+z}\sim\sum (\frac{2n}{z})^k $.
It is clear that the main contribution comes from $k\simeq 2n_0$
and $(2n_0)^{2n_0}\sim (2n_0)!$   exactly corresponds with the behavior
found above. It is in agreement with phenomenological analysis
\cite{Zakharov}, where it was assumed that the production
of a highly excited resonance might be responsible for
the large order behavior.

From the theoretical point of view we would expect
that this behavior is related, somehow, to purely
imaginary instantons (in order to provide the correct
$(-1)^k$ behavior). An additional arguments in favor of this point
will be discussed latter.

We would also like  to point out
that the numerical coefficient, which enters to the formula
(\ref{5a}) is follows:
\begin{equation}
\label{6a}
c_{2k}g^{2k}\sim  (\frac{g^2N_c}{Q^2})^{2k}  .
\end{equation} 
 At the same time, the perturbative $2k$-loop graph  gives a
different contribution $\sim (\frac{g^2N_c}{ \pi Q^2})^{2k} $
with an extra factor $1/\pi$   per each coupling constant.
This extra factor $\pi$  must be taken very seriously as it is a 
large parameter. We definitely know (from exact solution),
that the real scale of the problem is $m_0^2\pi^2$, and
not $m_0^2=g^2N_c/\pi$ as one would naively  
 expect from the
perturbative theory. This means that   vacuum condensates
which are determined by non-perturbative physics
come into the game. Even more, their contribution is much more important
than pure perturbative diagrams. Let us note, that the  lowest
vacuum condensates,   found exactly  in \cite{ARZ},
exhibit this additional factor $\pi$. Thus, the factorial growth
is related, somehow, to the nonperturbative physics.

To further investigate the nonperturbative
nature of the asymptotic series (in order to
support the    previous arguments)
let us, instead of correlation function 
(\ref{1a}),  consider the following difference  of correlators:
\begin{equation}
\label{delta}
i\int dx e^{iqx}\{  \langle 0|T \{ \bar{q}i \gamma_5q(x),
\bar{q}i \gamma_5q(0) \} |0 \rangle -
   \langle 0|T \{ \bar{q}q(x),
\bar{q}q(0) \} |0 \rangle \}
=  \Delta P(Q^2).
\end{equation}
One can argue that in the chiral limit $m_q\rightarrow 0$
the perturbative contribution to (\ref{delta}) iz zero. At the same time
dispersion relations lead  to the same
result: the coefficients of the OPE are factorially divergent.
This growth is related not to some perturbative diagrams,
but to   nonperturbative physics.  
We will present more arguments for this point of view in the next section.

 Finally we have    explicitly demonstrated that the OPE is 
an asymptotic series.
However we can not answer  the important question  of what kind
of operators are responsible for such behavior.
The reason for that is simple-- too many operators
contribute to the correlation function (\ref{1a})
and the corresponding classification problem
is quite involved. In the following we will consider
a special
heavy-light quark system, where such an identification
can be made. We find that some vacuum condensates exhibit a factorial
 growth. Exactly this
 fact is the   source of such an asymptotic behavior.

\subsection{ QCD coupled to adjoint fermions. Instantons.}
Now we   repeat the preceding  analysis for the much more
interesting model  
  of QCD with adjoint 
Majorana fermions   \cite{Bhanot,Klebanov,Kutasov}:
 \begin{equation}
S_{adj} = \int d^{2}x Tr \big[
-\frac{1}{4g^{2}} F_{\mu\nu}F^{\mu\nu} +
 i\bar{\Psi}\gamma^{\mu}D_{\mu}\Psi + m\bar{\Psi}\Psi\big]
\label{adjaction}
\end{equation}
 As is known, the most important difference with t'Hooft model
is that the bound states may contain, in general, {\it any} number of  
quanta.
In other  words,   pair creation is not suppressed even in the
large $N$ limit. The problem becomes more complicated, but much more
interesting, because the  pair creation imitates some physical
gluon effects.

We consider the following correlator analogous to (\ref{1a}):
\begin{equation}
\label{16}
i\int dx e^{iqx} \langle 0|T \{ \frac{1}{N_{c}}Tr\bar{\Psi} \Psi(x),
\frac{1}{N_{c}}Tr\bar{\Psi} \Psi(0) \} |0 \rangle = P_2(Q^2),
\end{equation}
where $\bar{\Psi}= {\Psi}^T\gamma_0$ and  the label $P_2$ shows
the number of partons in the  external source $\bar{\Psi} \Psi(x)$;
the factor $1/N_{c}$ is included  in  the definition
of the external current in order to make
the right hand side of the equation independent on $N$.
 In the large $Q^{2}$ limit the   leading contribution to correlation
 functions is given as before    by   
\begin{equation}
\label{17} 
P_2(Q^2\rightarrow\infty)=-\frac{2}{2\pi}\ln Q^2. 
\end{equation}
The additional
factor $2$ comes from two options in calculation of $Tr$ and
 related  to $Z_{2}$ symmetry mentioned in \cite{Kutasov}.

Now the problem arises. In t'Hooft model we definitely know
that only 2-particle bound states contribute to the corresponding
correlation function.
However, this  is not true for the model under consideration and
any states may contribute to $P_2$.  The {\it key} observation
is as follows: any pair creation  
(quantum loops which describe the virtual effects)
 is suppressed by a factor
$g^2N_{c}/Q^2$ because of dimensionality of the coupling constant
in two-dimensions (in a big contrast with
real 4-dimensional QCD)\footnote{Naively,
one could interpret such a result that the mixing
between  different number of partons,
is highly suppressed. Such a conclusion would
be in contradiction with numerical results
\cite{Bhanot}. However, as we argued in recent papers \cite{KogZhit}
the puzzle can be resolved by introducing
a nonzero value for vacuum condensate $\langle\bar{\Psi}  \Psi\rangle$.
Such a condensation does not  break any continuous symmetries.
Thus, no Goldstone Boson appears
as a consequence of the condensation.}.
 Besides that, the quark mass term
produces the analogous small factor $m_q^2/Q^2$ and can be neglected
as well. Thus, 
 information about highly excited states (which provides the $\log Q^2$
dependence) can be obtained exclusively from the analysis
of the correlation function at large $Q^2$. 
   In this case the analysis
is very similar to `t Hooft case, considered in the  previous section:
\begin{eqnarray}gin{eqnarray}
\label{18}
 \langle 0|\frac{1}{N_{c}}\bar{\Psi}  \Psi |n_1\rangle=
\sqrt{\frac{(m_0^2\pi^2)}{\pi}}f_{n_1} ,~~~n_1\gg 1,~~~~~
 n_1 \epsilon 2Z   \nonumber  \\
m_{n_1}^2=m_0^2\pi^2n_1,~~f_{n_1}^2=1  
,~~m_0^2=\frac{2g^2N_{c}}{\pi}. 
\end{eqnarray}  
The only difference  is the doubling of the  
strength of
the interaction $g^2\rightarrow 2g^2$, \cite{Klebanov} and the  
additional
degeneracy $Z_2$, mentioned above.
Now the formula for $P_2$ (\ref{17}) can be easily recovered:
\begin{equation}
\label{19} 
P_2(Q^2\rightarrow\infty)-P_2(0)=
\frac{2}{\pi}\sum_{n_1=0,2,4...}\frac{(m_0^2\pi^2)f_{n_1}^2}{m_{n_1}^2 
+Q^2}
 \rightarrow -\frac{2}{2\pi}\ln(Q^2),
\end{equation}
 As before, any correction to the asymptotic expression (\ref{7}),
such as  $ f_{n_1}^2=1+0(1/n_1),~
m_{n_1}^2=m_0^2\pi^2n_1+ 0(m_q) $ will produce   power corrections
$\sim 1/Q^2$ and they are not interesting at the moment.

Apparently, the formulae (\ref{17}-\ref{19})
are very similar to eqs. (\ref{2a}-\ref{4a}) which correspond to the 
't Hooft model. However, there is a big difference in interpretation
of these two cases: In the 't Hooft model we have exclusively
two- parton states (two bits, 
in terminology of refs.\cite{Klebanov},\cite{Bhanot}).
They saturate the dispersion relations. 

In the case  (\ref{17}-\ref{19})
we have much more  states with arbitrary number of partons.
As we explained in \cite{KogZhit} the mixing between the different 
numbers of partons is not suppresed because of 
$\langle\bar{\Psi}  \Psi\rangle$ -condensation. Effectively, however,
all these complex states contribute  to the correlation function 
(\ref{19}) in the same way as in the 't Hooft model.
In this case the integer number $n_1$ from (\ref{18})
should be interpreted as an excitation number
of 2- bits   in those states.
The matrix element $ f_{n_1}^2\simeq 1$ 
can be interpreted as a total probabilty
to find 2- bits   among  the complete set of the  mixed states.
The total number of states is increasing the mass increases.
Thus, the probability to find 2 -bits    in the given state is
decreasing correspondingly. However, the dispersion
relations (\ref{19}) tell us that the total probability $ f_{n_1}^2$
with the given excitation number $n_1$ remains the same.
 
We can repeat the previous analysis, which led us to the formula
(\ref{5a}) with  small changes. Instead of factor $B_{2k}$
in the expression (\ref{5a}) we will find  $$
B_{2k}\Rightarrow 2^{2k}B_{2k}$$ for the 
theory with adjoint matter. This difference was mentioned above
and is  related to the doubling of the strength of the interaction
$g^2\rightarrow 2g^2$.

How   could one interpret this result?
First of all, let us recall, that factorial behavior may  
occur for three different reasons: ultraviolet renormalons,
infrared renormalons and  instantons. Clearly, the 
  first two reasons can not   cause
for such  behavior in a field theory   with
the  dimensional coupling constant $g$. Thus
we expect
that some kind of classical solution should be responsible
for such behavior.
 If we accept the instanton hypothesis, then
from the very general arguments, one would expect
that the instanton 
contribution with action $S$ to 
the large  $k$-order coefficients $c_k$ is given by \cite{Large}:
\begin{equation}
\label{20}
c_k (g^2)^k\sim (k)! S^{-k} (g^2)^k
\end{equation}
In this case the factor $2$, mentioned above
has  the following interpretation: when we go from
the QCD with fundamental matter to the theory with adjoint
matter the instanton action is decreased by a factor of  $2$. In this case
the factor $ S^{-k}$ from the formula (\ref{20}) is exactly equal $2^k$.
It can be interpreted as the decreasing of instanton action by 
factor $2$.
Why the instanton with the action one-half is allowed in
the theory with adjoint matter and forbidden in the theory
with fundamental fermions ? This question has yet  to be answered.

Let us conclude this section by noting that from arguments given
above we expect that some classical, pure imaginary solution
(we call it instanton),  is responsible for the
factorial behavior found above. 

\section {Heavy-light quark system in $QCD_2$.}
\subsection {General remarks. }

As we mentioned in previous sections we are  not  able to identify the 
$n!$ behavior in the OPE (found from the spectrum) with some 
specific vacuum condensates. Such an identification can be done
if one considers the heavy-light quark system.
 In this case the problem is reduced to the analysis
 of vacuum expectation  value 
of the  Wilson line  $\langle W\rangle=
\langle\bar{q}(x)P\exp(ig\int_0^xA_{\mu}dx_{\mu})q(0)\rangle $.  
 Indeed, if we consider as in  \cite{Shuryak},\cite{Rad2}
 the correlation function
$\langle T\{ \bar{q}Q(x),\bar{Q}q(0)\}\rangle$, describing this system,
we end up (in the limit $M_Q\rightarrow\infty$) with the object
which is completely factorized (in accordance with 
HQET, see e.g. review \cite{Wise})
from the heavy quark: 
\begin{eqnarray}
\label{26} 
\langle T\{ \bar{q}Q(x),\bar{Q}q(0)\}\rangle
\sim  \langle\bar{q}(x)P\exp(ig\int_0^xA_{\mu}dx_{\mu})q(0)\rangle 
+{\em perturb.~ part} .
\end{eqnarray}
By definition, $\langle W\rangle$  in this formula 
is understood as the Taylor expansion:
\begin{eqnarray}
\label{27}
\langle W\rangle= \langle 0|\bar{q}(x) 
Pe^{ig\int_0^xA_{\mu}dx_{\mu}}q(0) |0\rangle  
=\sum_{n=0}^{n=\infty}\frac{1}{(2n)!}\langle
\bar{q}(x_{\mu}D_{\mu})^{2n}q\rangle  
 \end{eqnarray}
All nontrivial, large distance physics of
the system is hidden there.  
 Together with perturbative contributions  one should expect the
following behavior for this correlator\cite{Rad2}:
\begin{eqnarray}
\label{28} 
\langle T\{ \bar{q}Q(x),\bar{Q}q(0)\}\rangle
\sim   e^{-\Lambda \cdot x}.
\end{eqnarray}
 The perturbative terms,
proportional to $\sim(\frac{g^2N}{\pi})^n(x^2)^n$,
 contribute to (\ref{28})  as nonperturbative ones
due to the dimensional coupling constant $g$  in 2d,
thus, they interfere  with expansion (\ref{27}).
 As   was suggested
in the same context by Shifman 
\cite{Shifman}, one can get  rid of the perturbative terms by considering
 a special combination of scalar and pseudoscalar correlation functions
(analogous to (\ref{delta}) with the replacement
of a  light quark for a  heavy quark). The perturbative contribution vanishes 
in the chiral limit for the such combination and
we can study the pure nonperturbative physics.

Let us note that the general  connection (based   on
  dispersion relations)
between spectrum and vacuum condensates 
was considered  earlier\cite{Shifman}.
It was proven that the OPE is asymptotic series. 
Besides that,  for demonstration
 purposes,  it was suggested
  a specific  model for the spectral density
(linear trajectory) and it was found that  
the vacuum condensates 
(\ref{27}) get the form: $\langle\bar{q}(x_{\mu}D_{\mu})^{2n}q\rangle
 \sim (2n)!$.

In this section
we essentially follow the steps from  the paper
\cite{Shifman} with the only difference being  that
we start from 
theory defined as $QCD_2$ and derive the  $Q\bar{q}$ spectrum from this 
  Lagrangian. We find that we have not a linear spectrum
but rather $E_n\sim\sqrt{n}$ in $QCD_2$ for a heavy-light quark system. It gives 
a  different behavior for the vacuum condensates
$\langle\bar{q}(x_{\mu}D_{\mu})^{2n}q\rangle \sim n!$.
However, the main statement that the OPE is an asymptotic series
remains the same.

Before going  on, we would like to make the following remark.
In $QCD_2$ one can calculate
the appropriate vacuum condensates in the chiral limit 
from   first principles \cite{Chibisov}.  
Such a  calculation (which will be reviewed
 in the next section)
leads to   puzzling results. Roughly speaking, the results
of direct computation do not agree with indirect  calculations
based on the dispersion relations and spectrum.
  We formulate this puzzle as well as its resolution in the next section.
Anticipating the event, we would like to note here
that the origin of the puzzle is the
 factorial divergent coefficients
in the asymptotic series. If these expansions were convergent series,
we  would expect an  exact coincidence  of the results, based on two these
methods.
However, the analysis in field theory demonstrates 
 that this is not the case.

\subsection{ Spectrum $\bar{Q}q$ 
system in $QCD_2(N=\infty)$.}
As we have discussed in previous sections, if we knew
the spectrum of highly excited states we would calculate
(via dispersion relations )
the large order behavior for the corresponding 
correlation function. As we mentioned above, the heavy-light
quark system is very special in this sense, because
it allows us to identify the corresponding factorial behavior with
specific vacuum condensates. This is the main motivation
for the present section: find the spectrum
for highly excited states.
 Let us note that the heavy-light
quark system in this model
was considered previously
in ref.\cite{Ben}, but in a quite different context.

As we discussed in section 2,
the spectrum of highly excited states in $QCD_2$ is linear
(\ref{4}). This is certainly true, but only for the finite
parameter  $m_Q,  with ~n\rightarrow\infty$. We are now 
interested   in  a different limit, when $m_Q\rightarrow\infty$ first, and  
 $n\gg 1$ afterwards.
These limits do not commute and we have to start our analysis
from  exact original equation (\ref{2}).

In order to perform  the limit 
$m_Q\rightarrow\infty$ in `t Hooft equation (\ref{2}), let us make the following
change of variables:
\begin{eqnarray}
\label{29}
x=1-\frac{m_0}{m_Q}\alpha .~~~ 0\leq\alpha\leq\frac{m_Q}{m_0}=\infty .
\end{eqnarray}
Additionally we would like to rescale the wave function
and redefine the energy scale (from now on,
the counting of the  energy starts from 
$m_Q$) in the following way:
\begin{eqnarray}
\label{30}
m_n^2=(m_Q+E_n)^2\simeq m_Q^2+2m_QE_n ,~~\phi_n(x)\rightarrow
\sqrt{\frac{m_Q}{m_0}}\widetilde{\phi_n(\alpha)}.
\end{eqnarray}
Once  all these changes have been made
we arrive
to the following
equation, which replaces the `t Hooft equation,  for the  heavy-light quark
system in the  limit $m_Q\rightarrow\infty$.
\begin{eqnarray}
\label{31}
2E_n\widetilde{\phi_n(\alpha)}=m_0[\alpha+\frac{1}{\alpha}\frac{m_q^2-m_0^2}{m_0^2}]
\widetilde{\phi_n(\alpha)}
-m_0^2P\int_0^{\infty} d\beta\frac{\widetilde{\phi_n(\beta)}}{(\alpha-\beta)^2},
\end{eqnarray}
 The new set of wave functions in terms of new variables
is orthogonal and complete:
\begin{eqnarray}
\label{32}
\sum_n\widetilde{\phi_n(\alpha)}\widetilde{\phi_n(\beta)}
=\delta(\alpha-\beta),~~~
\int_0^{\infty}d\alpha\widetilde{\phi_n(\alpha)}\widetilde{\phi_m(\alpha)}
=\delta_{nm}
\end{eqnarray}

For the future analysis
we need not only the wave functions, but also
some physical matrix elements in terms of these wave functions.
It is convenient to separate the common factor, related to $m_Q$
and define the matrix elements in the following way
\footnote{We keep  the same notation $f_n$ for the corresponding
matrix elements. For light quark system they are defined in the different 
way (\ref{8}). We hope it will not confuse the reader.}:
\begin{eqnarray} 
\label{33}
 \langle 0|\bar{q}i\gamma_{5}Q |n\rangle
=\sqrt{\frac{N}{\pi}}\sqrt{m_0m_Q}f_n 
 ,~~f_n = \int_0^{\infty}d\alpha\widetilde{\phi_n(\alpha)}.
\end{eqnarray}
Using the parity relation \cite{Callan}, which in our notations
takes the form,
\begin{eqnarray}
\label{34}
\int_0^{\infty}d\alpha\widetilde{\phi_n(\alpha)}
=\frac{m_q}{m_0}\int_0^{\infty}d\alpha
\frac{\widetilde{\phi_n(\alpha)}}{\alpha}
\end{eqnarray}
one can show that the scalar matrix elements
have the same expression in terms of wave functions
as the pseudoscalar ones (\ref{33}).

Having these results in mind, 
and  using the standard technicques\cite{Hooft}- \cite{Brower1},
one can 
calculate \cite{ARZnew}the matrix elements $f_n$
and energies $E_n$ in the quasiclassical 
approximation for   $n\gg 1 $:
\begin{eqnarray}
\label{35}
E_n=2m_0\sqrt{\pi n}(1  +0(\frac{\log n}{n})),~~
f_n^2=\sqrt{\frac{\pi}{n}}(1 + 0(\frac{\log n}{n})).
\end{eqnarray}
This is the main result of this section.
It will be used in what follows for the 
calculation of large order behavior and 
high dimensional condensates.

Let us conclude this section with few remarks.
First of all, as  was expected, in the  
$m_Q\rightarrow\infty$ limit the equations (\ref{31}- \ref{35}),
do not depend on $m_Q$ (after an appropriate rescaling)
 in accordance with HQET
(see e.g. review \cite{Wise}).

Our second remark is the observation that
the chiral limit $m_q=0$ is very peculiar.
In particular, one can not take the limit 
$m_q=0$ in the identity (\ref{34}), 
because it clearly leads to a nonsensical result.
The reason for that is very simple. The edge region
$\alpha\sim m_q$ plays a very important role   in making 
 the theory (and this 
identity in particular)   selfconsistent. 
As a consequence,   one can not derive the boundary conditions
on wave function from the
 truncated equation  (\ref{31}) where the limit $m_Q\rightarrow \infty$ already
has been taken. In order to do so we need to come back to the original
`t Hooft eq.(\ref{2}).
We shall return to this point   in the  Conclusion.
We believe that the situation in four dimensional QCD
is quite similar in that the information about edge behavior
in the theory
can not be found from the truncated Lagrangian with the limit
$m_Q\rightarrow \infty$ already taken.
\subsection{High order condensates from duality and dispersion
relations.}

The starting point, as usual, is the correlation function
\begin{eqnarray}
\label{36} 
P(Q^2)=
i\int e^{iqx}dx  \langle 0| T\{ \bar{q}Q(x),\bar{Q}q(0)\}|0\rangle 
\end{eqnarray}
\begin{eqnarray}
\sim i\int e^{iqx}dx  \langle 0|\bar{q}(x)Pe^{ig\int_0^xA_{\mu}dx_{\mu}}q(0)|0\rangle 
+{\em perturb.~ part} . \nonumber
\end{eqnarray}
We follow \cite{Shuryak} and choose the external
momentum $q_0=m_Q-E,~~q_1=0$ very close to threshold.
 For positive  values of $E$
the correlation function can be written in the following way:
\begin{eqnarray}
\label{37}
P(E)= 
 \int_0^{\infty} e^{-Et}dt  \langle 0|\bar{q}(t)Pe^{ig\int_0^tA_{0}dt}q(0) |0\rangle 
+{\em perturb.~ part}.
\end{eqnarray}
Let us note that we could consider the difference
of two correlation functions (like (\ref{delta})).
 In that case the  perturbative contribution
to   (\ref{37}) would vanish. 

For large enough $E\gg m_0$ one can expand eq.(\ref{37}) in
$1/E$ \cite{Shifman}:
\begin{eqnarray}
\label{38}
P(E)=\frac{1}{E}[\langle\bar{q}q\rangle-\frac{1}{E^2}
\langle\bar{q}P_0^2q\rangle+
\frac{1}{E^4} \langle\bar{q}P_0^4q\rangle-
  ...] +perturb. part ,
\end{eqnarray}
where $P_0=iD_0$ is the time component of the momentum operator.

Our goal now is to substitute the asymptotic  expression (\ref{35})
for the matrix elements $f_n$ and energies $E_n$
into the dispersion relations
(analogous to (\ref{3a})). These will determine 
the higher order corrections  to the
correlation function as well as 
  the large $n$ behavior of the vacuum condensates
$\langle\bar{q}D^{2n}q\rangle$.

The appropriate dispersion relation states
that
\begin{eqnarray}
\label{39}
P(E)=\frac{N}{2\pi}m_0\sqrt{\pi}\sum_n\frac{f_n^2}{E+E_n}
\sim\frac{N}{\pi}\sum\frac{1}{\sqrt{n}(\sqrt{n}+\epsilon)}
\end{eqnarray}
where $\epsilon$ is external energy measured in $m_0\sqrt{\pi}$
units. 
In this formula we have taken into account 
 the following key observation
which we have used  earlier in the deriviation of  eq.(\ref{5a}):
the corrections  $\sim 1/n$ to  the asymptotic  behavior 
of the residues $f_n$ and energies $E_n$ (\ref{35})
might change the preasymptotic factor
for the  large order behavior. However, these corrections can not
change the main result-- the factorial growth of the coefficients found
below. This is the reason why we can not calculate
the corresponding coefficients {\it exactly}, but only
the leading factorial  factor. For the same reason we do not consider
the special difference of two correlation functions
(like (\ref{delta})), where perturbative contribution
is exactly canceled. In this case if we knew 
$f_n, E_n$ exactly, we would calculate the nonperturbative
condensates {\it exactly}! Unfortunately, this is not the case.
Thus, we ignore all  complications related to the separation
of pure nonperturbative contribution from the perturbative one.

Let us note that the sum in eq.(\ref{39})
is divergent at large $n$. This divergence is related
to the necessity of a  subtraction in the dispersion integral:
$P(E)\rightarrow P(E)-P(0)$. Besides that, at large energy
$E\gg m_0$  one can estimate the behavior $P(E\rightarrow
\infty)\sim \log E$,
which corresponds to the pure perturbative one loop
diagram.  These same features were present in our
analogous  previous formula
(\ref{5a}).

With these remarks in mind, our problem
is reduced to the  calculation of the coefficients
$c_k$  at large $k$  in the following  expansion:
\begin{eqnarray}
\label{40}
P(\epsilon)- subtractions  
\sim \sum_n\frac{1}{ \sqrt{n}+\epsilon}-  subtractions 
\end{eqnarray}
\begin{eqnarray}
\sim \log \epsilon +\sum_kc_k\frac{1}{\epsilon^k }. \nonumber
\end{eqnarray}
We note that the only difference with the previous formula (\ref{5a})
is   the dependence of the sum on $\sqrt{n}$ rather than $n$ itself.

The way to evaluate the coefficients $c_k$ 
is as follows.
We are going to use the standard idea (see, e.g. \cite{Prudnikov})
to present the sum in terms of the integral:
\begin{eqnarray}
\label{41}
\sum_{n=1}^{n=\infty}F(n)=\int_0^{\infty}\frac{f(x)}{e^x-1}dx,~~~
F(x)=\int_0^{\infty}f(x)e^{-xt}dt
\end{eqnarray}
However, in our case we have   $\sqrt{n} $
rather than    $n$ itself.
The corresponding generalization   is  known as well:
\begin{eqnarray}
\label{42}
\sum_{n=1}^{n=\infty}F[g(n)]=
\int_0^{\infty}\int_0^{\infty}\frac{h(x,y)f(y)}{e^x-1}dxdy,~~~
e^{-yg(p)}=\int_0^{\infty}h(x,y)e^{-xp}dx.
\end{eqnarray}
For our particular case  this formula gives the following
representation for the sum (\ref{40}):
\begin{eqnarray}
\label{43}
P(\epsilon)\sim\sum_n\frac{1}{ \sqrt{n+c
}+\epsilon}=\frac{1}{2\sqrt{\pi}}\int_0^{\infty}dx\int_0^{\infty}dy
\frac{ye^{-\frac{y^2}{4x}-cx-\epsilon y}}{x^{3/2}(e^x-1)}
\end{eqnarray}
where we have introduced  a constant $c$ for the future convenience
as an auxiliary parameter.
Such a parameter is actually present in 
original formulae (\ref{35}), however 
we believe that the main factorial dependence does not depend
on it.  Thus, after  differentiating with respect to $c$
  we put $c=0$ at the very end of calculation.
One more remark regarding formula (\ref{43}). Only  even $n$ should
be taken into account in this formula.  
However, by redefinition of parameters $c$ and  energy $\epsilon$,
the problem can be reduced to the same  integral.
 The result is an extra power dependence,
which is beyound our scope of    interest. Additionally, a
subtraction  which
should be made in this formula to get a convergent result,
has  no influence on $c_k$ at large $k$ (\ref{40}).

 The integration  over  $ x$ in the formula (\ref{43})
can be executed using the following expansion:
\begin{eqnarray}
\label{44}
\frac{x}{e^x-1}=\sum_{k=0}^{\infty}B_k\frac{x^k}{k!}
\end{eqnarray}
where $B_k$ are Bernoulli numbers.
Bearing in mind  that the appropriate integral from the
(\ref{43}) is known exactly
\begin{eqnarray}
\label{45}
\int_0^{\infty}\frac{e^{-\frac{y^2}{4x}-cx}}{\sqrt{x}}dx
=\sqrt{\frac{\pi}{c}}e^{-y\sqrt{c}},
\end{eqnarray}
and replacing $x^k$ from the formula (\ref{44})
by $(-1)^k(\frac{d}{dc})^k$,
we arrive to the following
expression for the sum (\ref{43}):
\begin{eqnarray}
\label{46}
P(\epsilon)\sim\frac{1}{2}\int_0^{\infty}dye^{-y\epsilon}y
\sum_{k=2}^{\infty}\frac{B_k}{k!}(-1)^k(\frac{d}{dc})^k
(\sqrt{\frac{1}{c}}e^{-y\sqrt{c}}).
\end{eqnarray}
 In this formula we ignore few first terms (proportional to $B_0, B_1$)
because they: a)do not contribute to large order
coefficients $c_k, k\gg 1$ , b)  they are  divergent and 
thus, require some subtractions discussed previously.

The key observation is as follows:
The nonperturbative part 
(which is our main interest) in the expansion (\ref{38})
is determined by odd powers of $1/(\epsilon)^{2m+1}$.
Such terms can be easily extracted from the formula
(\ref{46}) by expanding the exponent
\begin{eqnarray}
\label{47}
(\sqrt{\frac{1}{c}}e^{-y\sqrt{c}})=
\sum_l\frac{(-1)^ly^lc^{\frac{l-1}{2}}}{l!}
\end{eqnarray}
and executing the integration over $y$:
\begin{eqnarray}
\label{48}
\int_0^{\infty}dye^{-\epsilon y}y\frac{y^l}{l!}=
\frac{l+1}{\epsilon^{l+2}}.
\end{eqnarray}
It is now clear, that   only odd $l=2m-1$ terms
in the formula (\ref{47}) contribute to the 
coefficients related to nonperturbative part(\ref{38}). 
For small  parameter $c$ the coefficients $c_k$ can be easily
calculated \footnote{ As we mentioned before 
the factorial behavior does not depend on the particular
magnitude $c$. However, for $c=0$ the calculations
are much easier to present. Nonzero values of the  parameter $c$ might change
the preassymptotic behavior, which is beyound
  this method.}
by noting that for $l=2m-1$  the appropriate terms
from (\ref{47}) take the form 
$\frac{c^{\frac{l-1}{2}}}{l!}\sim \frac{c^{m-1}}{(2m-1)!}$.
The nonzero result in this case ( after differentiating 
$(\frac{d}{dc})^k $ and taking the  limit $c=0$ )
comes   exclusively from the term $k=m-1$ in eq.(\ref{46}).
Finally we arrive at the following
asymptotic expression for the odd coefficients
$c_k$ from the series (\ref{40}):
\begin{eqnarray}
\label{49}
c_{2k-1}\frac{1}{\epsilon^{2k-1}}\sim\frac{(-1)^kB_k}{k\epsilon^{2k-1}},
~~k\gg 1 .
\end{eqnarray}
Comparison with the original series (\ref{38}) suggests
that as dimension of the operator grows,  their
vacuum expectation values grow factorially,
\begin{eqnarray}
\label{50}
\langle\bar{q}P_0^{2n}q\rangle \sim\langle\bar{q}q\rangle (\pi m_0)^{2n}n!
\end{eqnarray}
From this formula one could naively think that
half of the VEV's in (\ref{49}) vanish because the odd
Bernoulli numbers are zero.
However, we think that this additional ``selection rule"
is accidental in its nature and thus, it should not be considered seriously.
We believe that this accidental vanishing of the coefficients  $c_k$
in the OPE
is related to our approximation (\ref{35}) for the
matrix elements $f_n$ and energies $E_n$. Thus, we expect,
that half of the VEV's which formally vanish in the leading approximation,
are actually not zero, but suppressed by a factor $\sim 1/n$, where
$n$ is dimension of an opeartor.
 
Having demonstrated the main result of this section, a few comments are in order.
 First of all, the OPE for $P(\epsilon)$ 
 is an asymptotic series as it must be in agreement with general
arguments \cite{Shifman}. 
Besides that, we expect the same behavior for analogous vacuum condensates
in four dimensional $QCD_4$ \cite{Zhit}.
Additionally, the scale for the vacuum condensates
is $m_0$. We shall see in the next section
that   different calculations of the same condensates   give somewhat different
results. We explain this puzzle later, but   note for now 
that $n!$ behavior in (\ref{50}) plays a crucial role 
in the explanation.

Finally, as we shall see, the $n!$ behavior has very general, 
  essentially kinematical origin in the large $N_c$ limit.
This  property  is related to so-called master field.
\section{High dimensional condensates from the theory. Puzzle.}

One can show \cite{Chibisov}, 
that the vacuum condensates which enter into the
formula (\ref{38}) at $N=\infty$ in the chiral limit $m_q\rightarrow 0$ in 
two dimensions  can be reduced
to a form which  contains 
  the field
strength tensor $ig\epsilon_{\mu\nu}
G_{\mu\nu} $ only. Indeed, 
the covariant derivatives
$\langle\bar{q} D^n\cdot D_{ \mu }D_{\mu}   q\rangle$
 placed at the very right and at the very left
(near the quark fields)
can be transformed into the operator  $ig\epsilon_{\mu\nu}
G_{\mu\nu} $ using the equation of motion
$D_{\mu}\gamma_{\mu}q=0$. To 
do the same thing with operators $D_{\mu}$
 which is placed somewhere in the middle of the expression,
we need to act, for example, on the right until the quark field is reached.
By doing so, step by step, we create many additional terms which
are either: commutator like 
$[D_{\mu},D_{\nu}]=-igG_{\mu\nu}$ which is the 
field strength operator  or
commutators like $\sim[D_{\lambda},\epsilon_{\mu\nu}
G_{\mu\nu}] $. Fortunately,   in two dimensions these terms are
related to creation of the  quark- antiquark fields
and we discard them in the chiral limit 
because they do not give $1/m_q$ enhancement, see formula (\ref{52})
 \footnote{
Of course this is not the case in four dimensions  
where $[D_{\mu}^2, G_{\lambda\sigma}]$ is an  independent 
operator which can not be reduced 
to   quark fields.}. 
We discuss the exact correspondence in the  Appendix, but for  now we  
 emphasize the existence of factor $n!$ in the corresponding 
formula:
\begin{eqnarray}
\label{51}
\langle\bar{q}(x_{\mu}D_{\mu})^{2n}q\rangle  \sim (x^2)^n n!
\langle  \bar{q}(gE)^nq  \rangle ,~~n\gg 1,~~~E\sim
\epsilon_{\lambda\sigma}G_{\lambda\sigma}.
 \end{eqnarray}

Thus, we end up with the   vacuum condensates 
$\langle  \bar{q}(gE)^nq  \rangle $ which are
expressed exclusively in terms of the field strength tensor.
Such vacuum condensates can be calculated  {\it exactly}
in the chiral limit\cite{Chibisov}.
The reason for this incredible simplification 
 is the observation that in   $QCD_2$  a gluon
is not a physical degree of freedom, but rather is a constrained
auxiliary field which can and should be expressed
in terms of the quark fields.
At the same time, in  the large 
$N$ limit,    the expectation
value of a product of any invariant operators reduces 
to their factorized values\cite{Witten1}.
 Exactly this feature of the large
$N$ limit (based on analysis of the so-called Master Field)
makes it possible to calculate the vacuum condensates exactly.
Our final expression takes the form \cite{Chibisov}:
\begin{eqnarray}
\label{52}
 \frac{1}{2^n} \langle  \bar{q}(ig\epsilon_{\lambda\sigma}
G_{\lambda\sigma}  \gamma_5)^nq \rangle 
 =(-\frac{g^2\langle\bar{q}q\rangle}{2m_q})^n\langle\bar{q}q\rangle 
 \end{eqnarray}
We interpret this expression as follows:
Each   insertion of an  additional factor
 proportional to field strength tensor $gE$, gives one and the same 
numerical factor (\ref{52}). This situation can be interpreted as
having a {\bf classical} master field \cite{Witten1}
which we insert in place of $gE$ in the vacuum condensates.
Because of its classical nature, it gives one and the same numerical factor.

Secondary, it is important to 
note that the vacuum condensate of an arbitrary local operator
can be reduced through the equation of motion and 
constraints to the fundamental quark condensate  (\ref{9}).

Finally, we must comment on the effective energy scale 
  which enters into the expression (\ref{52}): This  is not
$m_0^2$ as one could naively think, but rather
 $m_{eff}^2=m_0^3/m_q\gg m_0^2$. The obvious technical
reason for that is
related to the fact in the light cone gauge
(where the theory has been quantized)
$A_{-}=\frac{1}{\sqrt{2}}(A_0-A_1)=0$
  we have few constraints: the usual constraint in the gauge 
sector (Gauss law)\footnote{This formula explicitly demonstrates 
why   quark condensate appears in (\ref{52}) each time when we
insert an extra gluon field $E^{ab}$.}: 
\begin{eqnarray}
\label{constraint}
 \partial_{-}E^{ab}\sim g( q_+^{\dagger b} q_+^a-\frac{1}{N}
\delta^{ab}q_+^{\dagger c} q_+^c) ,
\end{eqnarray}
with right moving fermions $q_+$ considered as dynamical degrees of freedom.
   The left-moving fermions $q_-$
 are non-dynamical degrees of freedom in this gauge.  
The latter can be eliminated by the following  constraint:
\begin{eqnarray}
\label{53}
\frac{1}{\partial_-} q_+ ~ \sim \frac{1}{m_q}~q_- 
\end{eqnarray}
(for more details  see e.g. \cite{Lenz}).
This relation explicitly explains the origin
of the factor $1/m_q$  which is present in the formula (\ref{52})
(see \cite{Chibisov} for details).

Before   formulating  the puzzle, we would like to 
pause here in order to
   explain the  definition of the high dimensional
 vacuum condensates. As is known, they are 
  perturbatively  strongly
ultra violet (UV) divergent
objects. 

 As usual, all vacuum condensates should be understood 
in a sense that perturbative part is subtracted. The subtraction is organized by
 introducing of the so-called normalization parameter  $\mu$. In general, 
vacuum condensates do depend on this 
parameter $\mu$.  The gluon condensates of dimensions four, six,eight,...
in four dimensional $QCD$
are perfect examples where perturbative parts are divergent, put nonperturbative
 parts ( the remnants, which are left after subtraction)   are perfectly defined.
One could naively think that the instantons 
might spoil this picture, because
they give ultra -violet divergent contribution  to high order condensates.
However, we do not think this is the case and one can argue 
 that the definition we have  formulated remains untouched 
even with small size instanton effects taken
into account. 

The argument is as follows: We  should 
treat the  small size instantons and  the small size perturbative
contributions on the same basis. So, we should  subtract both  these
 divergences at the same time in order to get the so-called `` non-perturbative"
condensates which define the large distance physics.
Of course    small-distances physics does not disappear when
we do  such a subtraction. According to Wilson OPE
the corresponding 
 small distance contribution
(perturbative part and
small instanton contributions) should be taken
into account   separately.

In two dimensions this problem of course is much simpler. However,
the perturbative contributions to the condensates are divergent as well.
Nevertheless, the high dimensional condensates perfectly exist.
Indeed,  we have a   formula (\ref{52}), with 
  mixed condensates
expressed in terms of chiral condensate $\langle\bar{q}q\rangle$
where the latter  are    defined as the remnant after subtraction of
the perturbative contribution (for the
technical  details, see \cite{Mattis}).
Thus, our formula (\ref{52}) is  understood as a nonperturbative one, 
when we treat $\langle\bar{q}q\rangle$ as the nonperturbative chiral condensate.
Let us note, that  the gluon condensate in this model is finite
and can be calculated exactly \cite{ARZ} (see also the recent
paper \cite{Mattis} on this subject). In $QCD_2$ it can be expressed
in terms of the  chiral condensate.

 With this remark in mind, we are now ready to formulate the {\bf puzzle}.
The scale which enters into the OPE (\ref{38}) presumably 
is determined by the coefficients
$c_k$ from (\ref{40}). The leading contribution
to these  coefficients
can be expressed in terms of the vacuum condensates (\ref{52})
$$c_{2k+1}\sim (-\frac{g^2\langle\bar{q}q\rangle}{2m_q})^k\langle
\bar{q}q\rangle 
\sim (m_0^3/m_q)^k k!,~~           k\gg 1.$$
Thus, the characteristic scale of the problem
is $\frac{m_0^3}{m_q}$. 
We see an extra factor
$1/m_q$ in this scale.
It  has very clear origin (\ref{53}) and might be very big in the
chiral limit. At the same time, the characteristic scale which
can be found from the spectrum (\ref{5a},\ref{38}),
is much smaller and proportional to $m_0^2$. This is the puzzle.

The explanation of  this apparent paradox is as follows.
If our series (\ref{5a},\ref{38})  were convergent ones,
we would be in trouble, but  fortunately, our series are asymptotic ones.
Thus, in order to make sense 
to these series we have to define them and here we will use 
the standard 
    Borell summing prescription \cite{Large}
\footnote{We already mentioned in Introduction that the Borel summability
or its loss is not the crucial issue\cite{Fateev}. 
However, for the sake of definiteness
(and  for simplification) we assume in   general 
discussion which follows, that series  is Borel summable}.
Once this prescription has been accepted, we can write down
an expression which reproduces the asymptotic coefficients
for large number $k$ and at the same time is well defined
everywhere:
\begin{eqnarray}
\label{54}
\sum_k c_k \gamma^k\sim\int\frac{ d\gamma^{\prime}}{\gamma^{\prime}
(\gamma+\gamma^{\prime})}
e^{-\frac{1}{\gamma^{\prime}}},~~
c_k\sim (-1)^k k! \sim (-1)^k\int
\frac{d \gamma^{\prime}}{(\gamma^{\prime})^{k+2}}e^{-\frac{1}{\gamma^{\prime}}}.
\end{eqnarray}
We  did not specify the parameter $\gamma$ in this equation on 
purpose. Suppose, we have a large scale in the problem
determined by the vacuum condensates (\ref{52}).
In this case dimensionless parameter $\gamma$
has a large factor $1/m_q$:
\begin{eqnarray}
\label{55}
\gamma\sim\frac{(g^2N_c)^3}{m_q E^2}.
\end{eqnarray}
The prescription (\ref{54}) in this case states that the sum
of leading    terms $\sim (1/m_q)^k$
gives a zero contribution (in the chiral limit) 
\begin{eqnarray}
\label{56}
\sum_kc_k\gamma^k\sim\int\frac{ d\gamma^{\prime}}{\gamma^{\prime}
(\gamma+\gamma^{\prime})}
e^{-\frac{1}{\gamma^{\prime}}}\rightarrow 0
~(at~ m_q\rightarrow 0), ~~~\gamma\sim\frac{1}{m_q}
\end{eqnarray}
   to the physical correlation function!
We would like to note that the effect (\ref{56}) does not crucially
depend on the factorization properties for the 
condensates (\ref{52}) as neither on our assumption
of exact factorial dependence of the coefficients $c_k=k!(-)^k$ (\ref{54}).
Both of these effects  presumably lead
(apart to $k!$) to some mild $k$-dependence which can be easily
implemented into the formula (\ref{56}) by introducing some smooth function
$f(\gamma)$ whose moments
$$c_k=\frac{(-)^k}{k!}\int f(\gamma)\gamma^{-k-2}\exp(-\frac{1}{\gamma})
d\gamma \sim 1$$
exactly reproduce a $k$- dependence of the coefficients as 
well as of the condensates. If this function is mild enough,
it will not destroy the relation (\ref{56}), but might change some numerical
coefficients.

If we could calculate the vacuum condensates 
(\ref{52}) exactly (and not only the leading terms at $m_q\rightarrow 0$),
we would find that the  terms of order one give contribution
of order one to the correlation function.
  Thus, subleading terms play much more important role
in the final formula than the leading ones $\sim 1/m_q$. The origin of this
mystery of course is the factorial growth coefficients in the series.
This observation actually resolves the puzzle announced in the beginning
of the section.

We would like to make a few  more remarks  regarding this subject
in Conclusion which follows.
\section{Conclusion.}
We conclude this manuscript with the following lessons.

a). The OPE is asymptotic series.

b).We were lucky in a sense that in $QCD_2$ the scales 
for high dimensional condensates $\sim m_q^{-1}$ and  for the spectrum
 $\sim 1$ were parametrically different.
This was the reason why we noticed the difference very easily.
The lesson we can learn from this example is that
  numerically leading terms
  in the asymptotic expansion may
give somewhat negligible contribution into the final expression.
We believe that this is not a specific result for $2d$ physics, but rather
is a general property of a field theory associated with an asymptotic series.

c).Of course, we do not expect that
 in the real $QCD_4$ the vacuum condensates might 
exhibit some parameters similar to $\sim m_q^{-1}$. However, it might happen
that some subseries of 
 condensates possess a large numerical factor, let us call it $L\gg 1$.
In this case we would expect that the corresponding contribution
into the final formula will be suppressed, $L^{-1}$.
At the same time,  summing of a subseries 
of   terms which are 
  order of one,
 would lead to the result which is order of one.
  This observation raises the following question.

d). Suppose we have a  condensate $\langle O_k\rangle$
 of dimension $k$ which has 
both parts: the enhanced part, proportional to
 $L^k$  and the "regular" one of order $1^k$.
As we have learned, the subseries which has big factor like $L^k$
do not contribute much  to the final expression. At the same time
they are the main contributions into the  condensate with the given dimension $k$.

The moral is:  when we use   truncated expansions or approximate approaches
(like QCD sum rules), we inevitably study not
the actual condensates, but rather, some {\bf effective condensates }
\footnote{I thank
Michael Peskin for the discussion of this subject.
Actually, he raised the question of the  effective nature for the condensates
before this work was presented at the SLAC theory seminar.}.
One simple consequence of this is that the lattice calculation
of vacuum condensate might be different from QCD sum rules analysis. 

e).One may wonder what is the role of the  scale 
$1/m_q$ 
  in this model?
The answer is very peculiar.
Indeed this  scale can not be seen in spectrum however, from exact 
identities like  (\ref{34}) one can see that the edge region of order $m_q$
plays a very  important role in maintaining   selfconsistency 
of the theory.
The  calculation of the condensate
\cite{ARZ}, where the region $x\sim m_q$ gives whole answer is
another example of the same kind.
Exactly this infrared region determines the scale for the condensates
(\ref{52}), but not the scale for   integral characteristics like the spectrum
itself. See also remark after (\ref{35}) on the same subject.  

f). Since OPE is asymptotic series,  
it is good idea  to keep  only a few first terms  in the expansion
(like people do in 
the standard QCD sum rules approach, \cite{Shif}) and to stop at some 
point\footnote{In asymptotic series with coupling constant $\lambda$
one should stop with the number of terms of order $\lambda^{-1}$.
 In particular, in case of QCD sum rules, where $\lambda \sim 1/3\div 1/5$
 to be  determined by  the scale where
power corections are $20\div 30\% $, one can estimate the
maximum number of
terms in the expansion  is  about $3\div 5$.}.Any hopes to
 improve 
the standard QCD sum rules (like the idea advocated in 
\cite{Radyushkin})
by summing up of the certain subset of the power corrections, and ignoring
all the rest, might lead to the results which are much worse than 
the ones which follow from pre-improved version
of the approach. At least in $QCD_2$ such a procedure,
as we learned, gives a parametrically 
incorrect result. \\

We would like to make two more remarks
    which were  not our main subjects,  but
  look  like interesting byproducts worthy of    mention.\\

g).We argued that the $n!$ behavior found in $QCD_2$ 
is related to {\bf instantons}. Even more, the action
of the instanton in the theory with adjoint matter
is a  factor of  two less than in   theory with
  fundamental quarks. The explicit realization of
 such a solution is still lacking. 

h).We analyzed the $Q\bar{q}$ system in $QCD_2$. We found
that the edge behavior
of the system is very peculiar and can not be found from
the truncated theory, where limit $m_Q=\infty$ is already taken.
We believe that the same behavior is an inevitable feature
 of the real $QCD_4$.

\section{Acknowledgments.}
I would like to thank Michael Peskin          
    and other participants of 
 the SLAC theory seminar for   tough questions 
regarding this subject. I wish to thank Michael Shifman
for very useful criticism.
\section{Appendix.} 
 
The main goal of this Appendix is a demonstration 
of the factor $n!$ in formula (\ref{A})
in the chiral limit $m_q=0$. 
 
\begin{eqnarray}
\label{A}
\langle\bar{q}(x_{\mu}D_{\mu})^{2n}q\rangle  \sim (x^2)^n n!
\langle  \bar{q}(gE)^nq  \rangle ,~~n\gg 1, ~~~E\sim
\epsilon_{\lambda\sigma}G_{\lambda\sigma}.
 \end{eqnarray}
 We sketch  the idea only, ignoring for simplicity 
all normalization  factors.

We use the standard representation for $\gamma_{\mu}$ matrices
satisfying
\begin{eqnarray}
\label{A1}
\{\gamma_{\mu},\gamma_{\nu}\}=2g_{\mu\nu},
\gamma_0=\sigma_x,~\gamma_1=-i\sigma_y,~\gamma_5=\sigma_3,\\
\gamma_{\pm}=\gamma_0\pm\gamma_1,~~ D_{\pm}=D_0\pm D_1,~~
\gamma_-^2=0,~~\gamma_+^2=0,~~
\gamma_{\pm}\gamma_{\mp}\sim 1\mp\gamma_5  \nonumber
\end{eqnarray}

Dirac equations take the following form in the chiral limit
\begin{eqnarray}
\label{A2}
(\gamma_+D_-+\gamma_-D_+)q=0.
\end{eqnarray}
By multiplying these equations by $\gamma_{\pm}$, we arrive
to the following relations
\begin{eqnarray}
\label{A3}
(1\pm\gamma_5)D_{\pm}q=0,~~~\bar{q}D_{\mp}(1\pm\gamma_5)=0.
\end{eqnarray}

Let us present our original condensate in the following form
\begin{eqnarray}
\label{A4}
\langle\bar{q}(x_{\mu}D_{\mu})^{2n}q\rangle\sim
\langle\bar{q}(\frac{1+\gamma_5}{2}+\frac{1-\gamma_5}{2})
(x_{+}D_{-}+x_-D_+)^{2n}q\rangle .
\end{eqnarray}
From Lorentz invariance it is clear that a  nonzero
result $\sim (x_-x_+)^n$ comes from   
terms with equal numbers of $D_- ~and ~D_+$ with all possible permutations.
Our problem is how to count them. First, we pick up the projector
$(1+\gamma_5)$ from  the expression (\ref{A4}). According to 
(\ref{A3}) we should move $D_+$ to the right and $D_-$ to the left
(to reach the quark field)
  using 
 the following 
commutation relations
\begin{eqnarray}
\label{A5}
[D_+,D_-]\sim\epsilon_{\mu\nu}G_{\mu\nu}\sim E.
\end{eqnarray}
Here the field strength $E$ can be considered as 
constant (not operator), because  acting   the operators
$D_{\pm}$ on $E$ leads to the pair creation. 
We neglect  everywhere such terms in the limit $N\rightarrow\infty,
m_q\rightarrow 0$. The contribution   with projection
$(1-\gamma_5)$ gives exactly the same result after relabelling
$D_-\Leftrightarrow D_+$ and repeating the procedure described above.

Now one can see that these calculations are 
very similar (in the algebraic sense) to the 
 oscillator problem  with ladder  operators satisfying   the standard relations
\begin{eqnarray}
\label{A6}
[a,a^+]=1,~~a|0\rangle=0,~~ \langle 0|a^+=0.
\end{eqnarray}
Our problem is the calculation
of mean value of the operator $x^{2n}\sim(a+a^+)^{2n}$ for the ground state:
\begin{eqnarray}
\label{A7}
 \langle 0|x^{2n}|0\rangle \sim \langle 0|(a+a^+)^{2n}|0\rangle \sim \Gamma(n+\frac{1}{2})
\sim n!,~~n\gg 1.
\end{eqnarray}
This concludes our explanation of the factor $n!$ 
(\ref{51}) which we heavily used 
in our previous discussions.

Let us repeat again that all factors from (\ref{51}, \ref{52}) have clear meaning
and can be explained without detailed  calculations.
Three steps are involved:
1.The transition from the operator 
$\langle\bar{q}(x_{\mu}D_{\mu})^{2n}q\rangle $ to the operator 
$\langle  \bar{q}(gE)^nq  \rangle $ with a factor $n!$ was explained in this
appendix.
2.The idea of Master Field predicts that each 
insertion of $E$   into the expression for the
operator  $\langle  \bar{q}(gE)^nq  \rangle $
gives one and the same constant $M_{eff}^n$.
3.The constraints (\ref{constraint},\ref{53}) explicitly demonstrate
that $M_{eff}^2\sim m_q^{-1}$.

\pagebreak


\begin{thebibliography}{xx}
\bibitem{Hooft} G.'t Hooft, Nucl.Phys. {\bf B75},(1974),461.
\bibitem{Callan} C.G.Callan, N.Coote and D.J.Gross,
Phys.Rev.{\bf D13},(1976),1649.
\bibitem{Einh}M.B.Einhorn, Phys.Rev.{\bf D14},(1976),3451.
\bibitem{Einhorn} M.B.Einhorn,S.Nussinov and E.Rabinovici,
Phys.Rev.{\bf D15},(1977),2282.
\bibitem{Brower}R. Brower,J.Ellis,M.Schmidt and J.Weis,
Nucl.Phys. {\bf B128},(1977),131; Nucl.Phys. {\bf B128},(1977),175..
\bibitem{Brower1} R.Brower, W.Spence and J.Weis,
Phys.Rev.{\bf D19},(1979),3024.
\bibitem{Kent} K.Hornbostel,
S.Brodsky and H.Pauli, Phys.Rev.{\bf D41},(1990),3814.
 \bibitem{Klebanov}S.Dalley and I.Klebanov, Phys.Rev.{\bf D47},(1993),2517. 
\bibitem{Kutasov} D.Kutasov, Nucl.Phys. {\bf B414},(1994),33; hep-th/9501024.
\bibitem{Fateev} V.A.Fateev, V.A.Kazakov, P.B.Wiegmann,
Principle Chiral Field at Large N.,ENS-94/07, March 1994.
\bibitem{Zakharov} V.I. Zakharov,
Nucl.Phys. {\bf B377}, (1992), 501;
  Nucl.Phys. {\bf B385}, (1992), 452. 
\bibitem{ARZ} A.Zhitnitsky, Phys.Lett.{\bf B165},(1985),405;
Sov.J.Nucl.Phys.{\bf 43}, (1986), 999;  Sov.J.Nucl.Phys.{\bf 44},(1986),139.
\bibitem{Ming1}Ming Li, Phys.Rev.{\bf D34},(1986),3888.
\bibitem{Ming2}Ming Li et el. J.Phys. {\bf G13},(1987),915.
\bibitem{Lenz}F.Lenz et al.Ann.Phys. {\bf 208},(1991),1.
\bibitem{Mattis}M.Burkardt, hep-ph/9409333.
\bibitem{Coleman}S.Coleman, Commun.Math.Phys.,{\bf 31},(1973),259. 
139.
 \bibitem{Witten}E.Witten, Nucl.Phys. {\bf B145},(1978),110.
\bibitem{BKT}V.Berezinski, JETP,{\bf 32},(1971),493;\\
J.Kosterlitz and D.Thouless, J.Phys.C {\bf 6},(1973),1181.
\bibitem{Bhanot}G.Bhanot, K.Demeterfi and I.Klebanov,
 Phys.Rev.{\bf D48},(1993),4980; Nucl.Phys. {\bf B418},(1994),15.
\bibitem{Shifman} M.A.Shifman, Theory of Pre-Asymptotic Effects in weak
Inclusive Decays, TPI-MINN-94/17-T, Talk at the Workshop, Minneapolis, 
hep-ph 9405246; Talk at PASCOS,Baltimore,1995,hep-ph/9505289.
\bibitem{KogZhit}  I. Kogan and A.Zhitnitsky
Two dimensional QCD with adjoint matter: What does it teach us?
hep-ph/9509322, to apper in Nuclear Physics B.
\bibitem{Large}Current Physics-Sources and Comments, vol.7,
"Large Order Behavior of Perturbative Theory",eds J.C.Gillou and 
J.Zinn-Justin,1990.
  \bibitem{Shuryak}E.Shuryak, Nucl.Phys. {\bf B198},(1982),83;
Nucl.Phys. {\bf B203},(1982),116.
\bibitem{Rad2}A.Radyushkin,Phys.Lett.
{\bf B271}, (1991),218.
\bibitem{Wise} N.Isgur and M.Wise , in ``B decays", Ed. S.Stone,
Word Scientific,1992.
\bibitem{Zhit} A. Zhitnitsky, Phys.Lett.{\bf B357},(1995),211;\\
B.Chibisov and A.Zhitnitsky,  
Phys.Rev.{\bf D52},(1995),5273.
\bibitem{Chibisov}B.Chibisov and A.Zhitnitsky,  
  Phys.Lett.{\bf B362},(1995),105.
\bibitem{Ben}B.Grinstein and P.Mende, Nucl.Phys.{\bf B425},(1994),451. 
  \bibitem{Prudnikov}Prudnikov, et al, Integrals and Series;
 I.Gradshtein and I.Ryzhik {\it Tables of Integrals, Series and
Products }, Academic Press, New-York (1965)
\bibitem{Witten1}E.Witten, In Recent Developments in Gauge Theories,
eds. G.'t Hooft et.al. Plenum Press, 1980.
\bibitem{ARZnew}A.Zhitnitsky,   in preparation, 1996.    
 \bibitem{Shif}
M.A.Shifman, A.I.Vainshtein and V.I.Zakharov 
   Nucl.Phys.  {\bf B147}, (1979)385,448,519. \\
M.A.Shifman, Vacuum Structure and QCD Sum Rules, 
North-Holland,1992.
\bibitem{Radyushkin}S.Mikhailov and A.Radyushkin, JETP Lett. {\bf 43},(1986),712;
Phys.Rev.{\bf D 45}(1992),1754.
     \end{thebibliography}
\end{document}